%% file: LFVHiggs_paper.tex
\documentclass[cernpreprint,texlive=2011,txfonts,UKenglish,biblatex=false,texmf]{atlasdoc} 
\pdfoutput=1

\usepackage{atlaspackage}

\usepackage{atlasphysics}

\usepackage{mathrsfs}
\usepackage{cite}
\usepackage{float}
\usepackage{lineno}
\usepackage{multicol}
\usepackage{multirow}
\usepackage{rotating}
\usepackage{morefloats}
\usepackage{rotating}
\usepackage{axodraw}
\usepackage{authblk}
\usepackage{tabulary}
\usepackage{longtable}
\usepackage{ amssymb }
\usepackage{footnote}
\usepackage{lscape}
\usepackage{graphicx} 
\DeclareGraphicsExtensions{.pdf,.eps}
\usepackage{rotating} 
\usepackage{slashbox} 

\usepackage{LFVHiggs_paper-defs}

\input{LFVHiggs_paper-metadata}
\hypersetup{pdftitle={ATLAS draft},pdfauthor={The ATLAS Collaboration}}

\begin{document}

\maketitle



\input paper_text

\section*{Acknowledgements}

\input{Acknowledgements}



\bibliographystyle{atlasBibStyleWithTitle}
\bibliography{bibliography,ATLAS}
\newpage
\input{atlas_authlist}

\end{document}

%% file: LFVHiggs_paper-metadata.tex
\AtlasTitle{Search for lepton--flavour--violating $H\to\mu\tau$ decays of the Higgs boson with the ATLAS detector}

\author{The ATLAS Collaboration}

\PreprintIdNumber{CERN-PH-EP-2015-184}

\AtlasJournal{JHEP}

\AtlasAbstract{%
A direct search for lepton--flavour--violating (LFV) $H\to\mu\tau$ decays of the 
recently discovered Higgs boson with the ATLAS detector at the LHC is presented. The
analysis is performed in the $H\to\mu\tau_{\mathrm{had}}$ channel, where $\tau_{\mathrm{had}}$ 
is a hadronically decaying $\tau$--lepton. The search is based on the data
sample of proton--proton collisions collected by the ATLAS experiment corresponding to
an integrated luminosity of 20.3~\ifb\ at a centre--of--mass energy of $\sqrt{s}=8$~TeV.
No statistically significant excess of data over the predicted background is observed.
The observed (expected) 95\% confidence--level upper limit on the branching
fraction, Br($H\to\mu\tau$), is 1.85\% (1.24\%). 
}

%% file: paper_text.tex
\section{Introduction}
\label{sec:introduction}

The observation of the Higgs boson~\cite{ATLAS_disc,CMS_disc} with a mass of about 125~GeV~\cite{Aad:2015zhl} by the
ATLAS and CMS experiments is a great success
of the Large Hadron Collider (LHC) physics program at CERN. The next 
important step in this program are searches for signs of new physics beyond 
the Standard Model (SM) and detailed studies of the Higgs boson properties. 
Direct evidence for physics beyond the SM could be indicated via
lepton--flavour--violating (LFV) Higgs boson decays.
If the SM is replaced with an effective field theory, which has a single Higgs boson and 
is required to be renormalizable only to a finite mass scale, then LFV couplings may 
be introduced~\cite{Harnik:2012pb}. LFV decays can also occur naturally in 
models with more than one Higgs doublet~\cite{Bjorken:1977vt,DiazCruz:1999xe,Han:2000jz,Arhrib:2012ax}, 
composite Higgs models~\cite{Agashe:2009di,Azatov:2009na}, models with flavour 
symmetries~\cite{Ishimori:2010au}, Randall--Sundrum models~\cite{Perez:2008ee} 
and many others~\cite{Blanke:2008zb,Giudice:2008uua,AguilarSaavedra:2009mx,Albrecht:2009xr,Goudelis:2011un,McKeen:2012av}. 

There are three possibilities for LFV effects mediated via virtual Higgs bosons: 
$\mu$--$e$, $\tau$--$\mu$, and $\tau$--$e$ transitions. Indirect 
experimental constraints are reviewed and translated into constraints on 
Br($H\to e\mu,\mu\tau,e\tau$) in recent papers~\cite{Harnik:2012pb,Blankenburg:2012ex}. 
Searches for $\mu\to e\gamma$~\cite{Adam:2013mnn} place a
very stringent constraint on $H\to e\mu$ decays: Br($H\to e\mu$)~$<$~O($10^{-8}$)~\cite{Harnik:2012pb,Blankenburg:2012ex}. 
The indirect constraints on $H\to\mu\tau,e\tau$ decays mostly come from
searches for $\tau\to\mu\gamma,e\gamma$~\cite{Aubert:2009ag,Kanemura:2005hr,Davidson:2010xv} 
or other rare $\tau$--lepton decays~\cite{Celis:2013xja}, as well as
from measurements of the anomalous magnetic moment of the muon and the electron~\cite{Agashe:2014kda} and are much less stringent: 
Br($H\to\mu\tau,e\tau$)~$<$~O(10$\%$)~\cite{Harnik:2012pb,Blankenburg:2012ex}. A relatively large Br($H\to\mu\tau$) 
can be achieved without any particular tuning of the effective couplings, while a large 
Br($H\to e\tau$) is possible only at the cost of some fine--tuning of the corresponding 
couplings~\cite{Blankenburg:2012ex}. It is also important to note that the presence 
of a $H\to\mu\tau$ signal would essentially exclude the presence of a $H\to e\tau$ signal, 
and vice versa, at an experimentally observable level at the LHC due to strong experimental 
bounds on $\mu\to e\gamma$ decays~\cite{Blankenburg:2012ex}. 
The CMS Collaboration has recently performed 
the first direct search for LFV $H\to\mu\tau$ decays~\cite{Khachatryan:1994783} and 
reported a slight excess (2.4 standard deviations) of data over the predicted background. 
Their results give a 1.57$\%$ upper limit on Br($H\to\mu\tau$) at the 95$\%$ confidence level (CL).

This paper presents a search for lepton--flavour--violating $H\to\mu\tau$ decays 
of the recently discovered Higgs boson with one muon and one hadronically decaying 
$\tau$--lepton ($\thad$) in the final state. 
The analysis is based on the data sample of $pp$ collisions which was collected at a 
centre--of--mass energy of $\sqrt{s}=8$~TeV and corresponds to an integrated luminosity of 20.3~\ifb.

\section{The ATLAS detector and object reconstruction}
\label{sec:object_reco}

The ATLAS detector\footnote{ATLAS uses a right-handed coordinate system with its origin at 
the nominal interaction point (IP) in the centre of the detector and the $z$-axis along the 
beam pipe. The $x$-axis points from the IP to the centre of the LHC ring, and the $y$-axis 
points upward. Cylindrical coordinates $(r,\phi)$ are used in the transverse plane, $\phi$ 
being the azimuthal angle around the beam pipe. The pseudorapidity is defined in terms of 
the polar angle $\theta$ as $\eta=-\ln\tan(\theta/2)$.} is described in detail in
ref.~\cite{Aad:2008zzm}. ATLAS consists of an inner tracking detector (ID) covering the range 
$|\eta|<2.5$, surrounded by a superconducting solenoid providing a 2~T magnetic field, 
electromagnetic ($|\eta|<3.2$) and hadronic calorimeters ($|\eta|<4.9$) and a muon 
spectrometer (MS) ($|\eta|<2.7$) with a toroidal magnetic field. 

The signature of LFV $H\to\mu\tau$ decays used in this search is characterised by the presence
of an energetic muon, originating directly from a Higgs boson decay and carrying roughly half of its 
energy, and the hadronic decay products of a $\tau$--lepton. The data were collected with a 
single--muon trigger with a transverse momentum, $\pt=p\sin{\theta}$, threshold of $\pt=24$~GeV. 
The $H\to\tau\tau$ and the LFV $H\to\mu\tau$ signatures with a muon and \thad in the final 
state share many common features.
Therefore, the object definitions and data quality cuts used in this
analysis are the same as those in the recently published ATLAS search for $H\to\tau\tau$ 
decays~\cite{Aad:2015vsa}. A brief description of the object definitions is provided below.

Muon candidates are reconstructed using an algorithm~\cite{Aad:2014rra} that combines 
information from the ID and the MS. Muon quality criteria such as inner detector 
hit requirements are applied to achieve a precise measurement of the muon momentum and 
to reduce the misidentification rate. Muons are required to have 
$\pt>$10~GeV and to be within $|\eta|<2.5$. Typical reconstruction and identification
efficiencies for muons satisfying these selection criteria are above 95\%~\cite{Aad:2014rra}. 
Exactly one identified muon is required in this analysis. 

Electron candidates are reconstructed from energy clusters in the electromagnetic calorimeters 
matched to tracks in the ID. They are required to have a transverse energy, $\et = E\sin{\theta}$, 
greater than 15~GeV, to be within the pseudorapidity range $|\eta|<2.47$, and to satisfy the 
{\it medium} shower shape and track selection criteria defined in ref.~\cite{Aad:2014fxa}. Candidates 
found in the transition region between the end--cap and barrel calorimeters ($1.37<|\eta|<1.52$) 
are not considered. Typical reconstruction and identification efficiencies for electrons satisfying 
these selection criteria range between 80\% and 90\% depending on $\et$ and $\eta$. Since electrons 
do not appear in the $H\to\mu+\thad$ decay mode, events with identified electrons are rejected. 

Jets are reconstructed using the anti--$k_{\mathrm{t}}$ jet clustering algorithm~\cite{Cacciari:2008gp} with a radius
parameter $R=0.4$, taking clusters of calorimeter cells with deposited energy as inputs. Fully calibrated 
jets~\cite{Aad:2012vm} are required to be reconstructed in the range $|\eta|<4.5$ and to have $\pt>30$~GeV. 
To reduce the contamination of jets by additional interactions in the same or neighbouring bunch crossings 
(pile--up), tracks originating from the primary vertex must contribute a large fraction of the jet $\pt$ 
when summing the scalar $\pt$ of all tracks in the jet. The primary vertex is chosen as the proton--proton 
collision vertex candidate with the highest sum of the squared transverse momenta of all associated tracks.
This jet vertex fraction is required to be at least 50\% for jets with $|\eta|<$2.4 and $\pt<$50~GeV 
(no cut is applied to jets with $\pt>50$~GeV). Jets with no associated tracks are retained. In the 
pseudorapidity range $|\eta|<2.5$, jets containing $b$--hadrons ($b$--jets) are selected using a tagging 
algorithm~\cite{ATLAS-CONF-2014-046}, which has an efficiency of $\sim$70\% for 
$b$--jets in $t\bar{t}$ events. The corresponding light--flavour jet misidentification probability 
is 0.1--1\%, depending on the $\pt$ and $\eta$ of the jet. Only a very small fraction of signal events have 
$b$--jets, therefore events with identified $b$--jets are vetoed in the selection of signal events.

Hadronically decaying $\tau$--leptons are identified by means of a multivariate analysis technique~\cite{Aad:2014rga}
based on boosted decision trees, which exploits information about ID tracks and clusters in 
the electromagnetic and hadronic calorimeters. The $\thad$ candidates are required to have charge 
$q_{\tau}=\pm 1$ in units of electron charge, and must be 1-- or 3--track (1-- or 3--prong) candidates. 
Events with exactly one $\thad$ candidate satisfying the medium identification criteria~\cite{Aad:2014rga} 
with $\pt>$20~GeV and $|\eta|<$2.47 are considered in this analysis. The identification efficiency for 
$\thad$ candidates satisfying these requirements is 55--60\%. Dedicated criteria~\cite{Aad:2014rga} to 
separate $\thad$ candidates from misidentified electrons are also applied, with a selection efficiency 
for true $\thad$ decays of 95\%. To reduce the contamination due to backgrounds where a muon fakes a $\thad$ 
signature, events where an identified muon with $\pt(\mu)>4$~GeV overlaps with an identified $\thad$ 
are rejected~\cite{Aad:2015vsa}. The probability to misidentify a jet with $\pt>20$~GeV as a $\thad$ 
candidate is typically 1--2\%.       

The missing transverse momentum (with magnitude \met) is reconstructed using the energy deposits 
in calorimeter cells calibrated according to the reconstructed physics objects ($e$, $\gamma$, 
$\thad$, jets and $\mu$) with which they are associated~\cite{Aad:2012re}. 
The energy from calorimeter cells not associated with any physics object is included in the $\met$ 
calculation. It is scaled by the ratio of the scalar sum of $\pt$ of tracks which originate from the 
primary vertex but are not matched to any objects and the scalar sum of $\pt$ of all tracks in the 
event which are not matched to objects. The scaling procedure achieves a more accurate reconstruction 
of $\met$ in high pile--up conditions. In this search, $\met$ is a signature of neutrinos, and it is used to 
select and reconstruct signal events, as described below.   

\section{Event selection and categorization}
\label{sec:selection}

Signal $H\to\mu\tau$ events in the $\mu\thad$ final state are characterised by the presence of an energetic 
muon and a $\thad$ of opposite charge as well as moderate $\met$, which tends to be aligned with the $\thad$ 
direction. Backgrounds for this signature can be broadly classified into two major categories:
\begin{itemize}
\item Events with true muon and $\thad$ signatures, dominated by irreducible $Z/\gamma^{*}\to\tau\tau$ production 
with some contributions from the $VV\to\mu\tau+X$ (where $V=W,Z$), $t\bar{t}$, single--top and SM 
$H\to\tau\tau$ production processes; these events exhibit a very strong charge correlation between 
the muon and the $\thad$; therefore, the expected number of events with opposite--sign (OS)
charges ($N_{\mathrm{OS}}$) is much larger than the number of events with same--sign (SS) charges ($N_{\mathrm{SS}}$). 
\item Events with a fake $\thad$ signature, dominated by $W$+jets events with some contribution 
from  multi--jet (many multi--jet background events have genuine muons from semileptonic 
decays of heavy--flavour hadrons), diboson ($VV$), $t\bar{t}$ and single--top events with some charge 
asymmetry $N_{\mathrm{OS}}>N_{\mathrm{SS}}$; $Z\to\mu\mu$+jets events, where a $\thad$ signature can be faked by either 
a jet (no charge correlation) or a muon (strong charge correlation), also contribute to this category. 
\end{itemize}

Events with a fake $\thad$ tend to have a much softer $\pt(\thad)$ spectrum and a larger angular separation 
between the $\thad$ and $\met$ directions. These properties are exploited to suppress such backgrounds and define 
signal and control regions. Events with exactly one muon and exactly one $\thad$ with $\pt(\mu)>26$~GeV, 
$\pt(\thad)>45$~GeV and $|\eta(\mu)-\eta(\thad)|<2$ form a baseline sample. The $|\eta(\mu)-\eta(\thad)|$ 
cut has $\sim$99\% efficiency for signal and rejects a considerable fraction of multi--jet and $W$+jets events. 
At this stage of the event selection, the identified muon is also required to be isolated~\cite{Aad:2015vsa} 
in the calorimeters and in the tracking detector in order to reduce contamination from the multi--jet background. Two signal 
regions are defined using the transverse mass, $\mT$,\footnote{$\mT=\sqrt{2\pt^{\ell}\met(1-\cos{\Delta\phi})}$, 
where $\ell=\mu$,~$\thad$ and $\Delta\phi$ is the azimuthal separation between the directions of the lepton 
($\mu$ or $\thad$) and $\met$ vectors.} of the $\mu$--$\met$ and $\thad$--$\met$ systems: OS events with 
$\mT(\mu,\met)>40$~GeV and $\mT(\thad,\met)<30$~GeV form the signal region--1 (SR1), while OS events 
with $\mT(\mu,\met)<40$~GeV and $\mT(\thad,\met)<60$~GeV form the signal region--2 (SR2). The dominant 
backgrounds in SR1 and SR2 are $W$+jets and $Z/\gamma^{*}\to\tau\tau$ events, respectively. The modelling 
of the $W$+jets background is checked in a dedicated control region (WCR) formed by events with 
$\mT(\mu,\met)>60$~GeV and $\mT(\thad,\met)>40$~GeV. The choice of the $\mT$ cuts to define SR1, SR2 and WCR 
is motivated by correlations between $\mT(\mu,\met)$ and $\mT(\thad,\met)$ in $H\to\mu\tau$ signal and major 
background ($W$+jets and $Z/\gamma^{*}\to\tau\tau$) events, as illustrated in figure~\ref{fig:mt_2Dplots}. 
No events with identified $b$--jets are allowed in SR1, SR2 and WCR. The modelling of the $t\bar{t}$ and 
single--top backgrounds is checked in a dedicated control region (TCR), formed by events that satisfy the 
baseline selection and have at least two jets, with at least one being $b$--tagged. Table~\ref{tab:cuts} 
provides a summary of the event selection cuts used to define the signal and control regions. 

\begin{figure}
\centering
\includegraphics[width=0.45\textwidth]{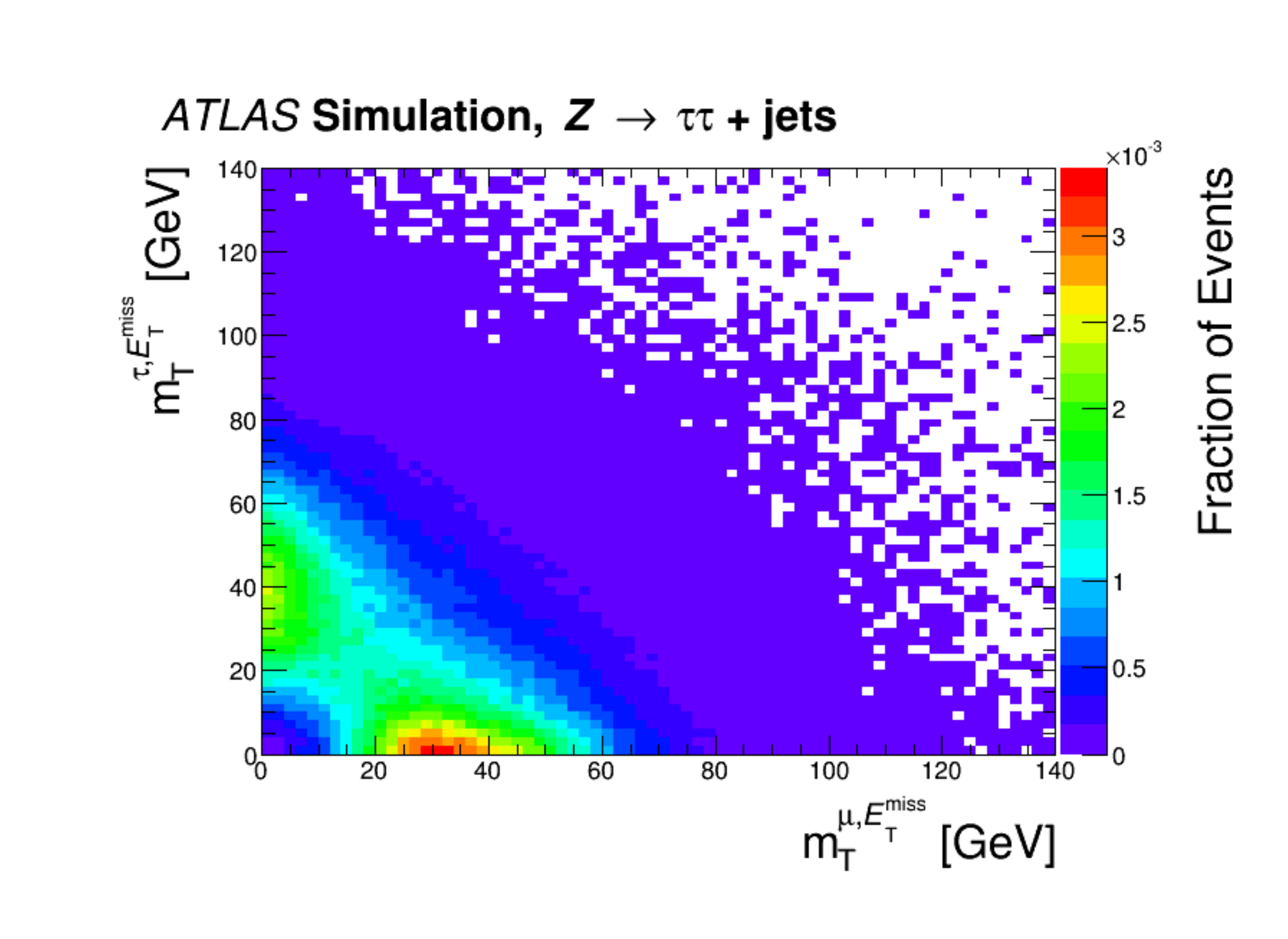}
\includegraphics[width=0.45\textwidth]{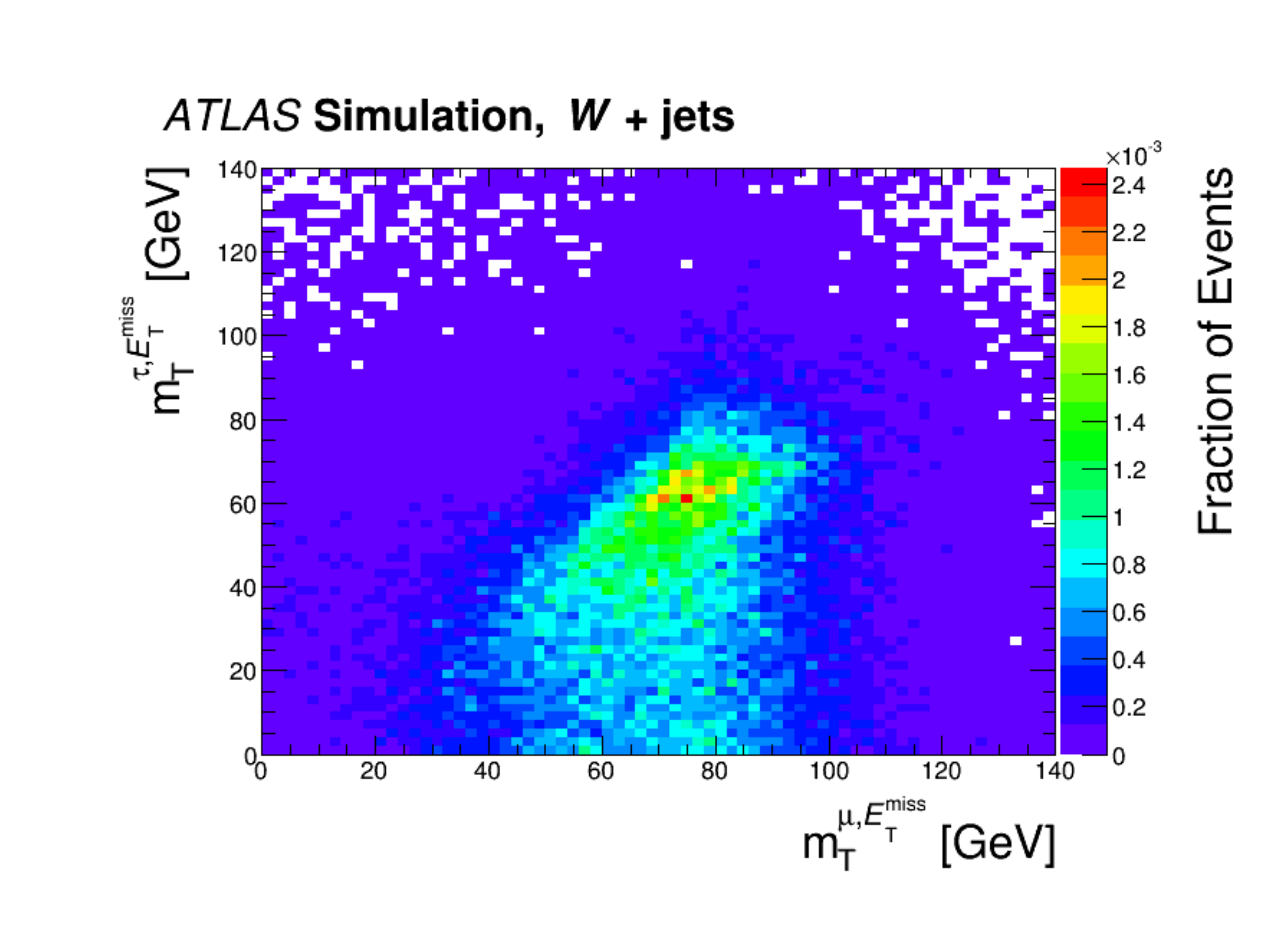}
\includegraphics[width=0.45\textwidth]{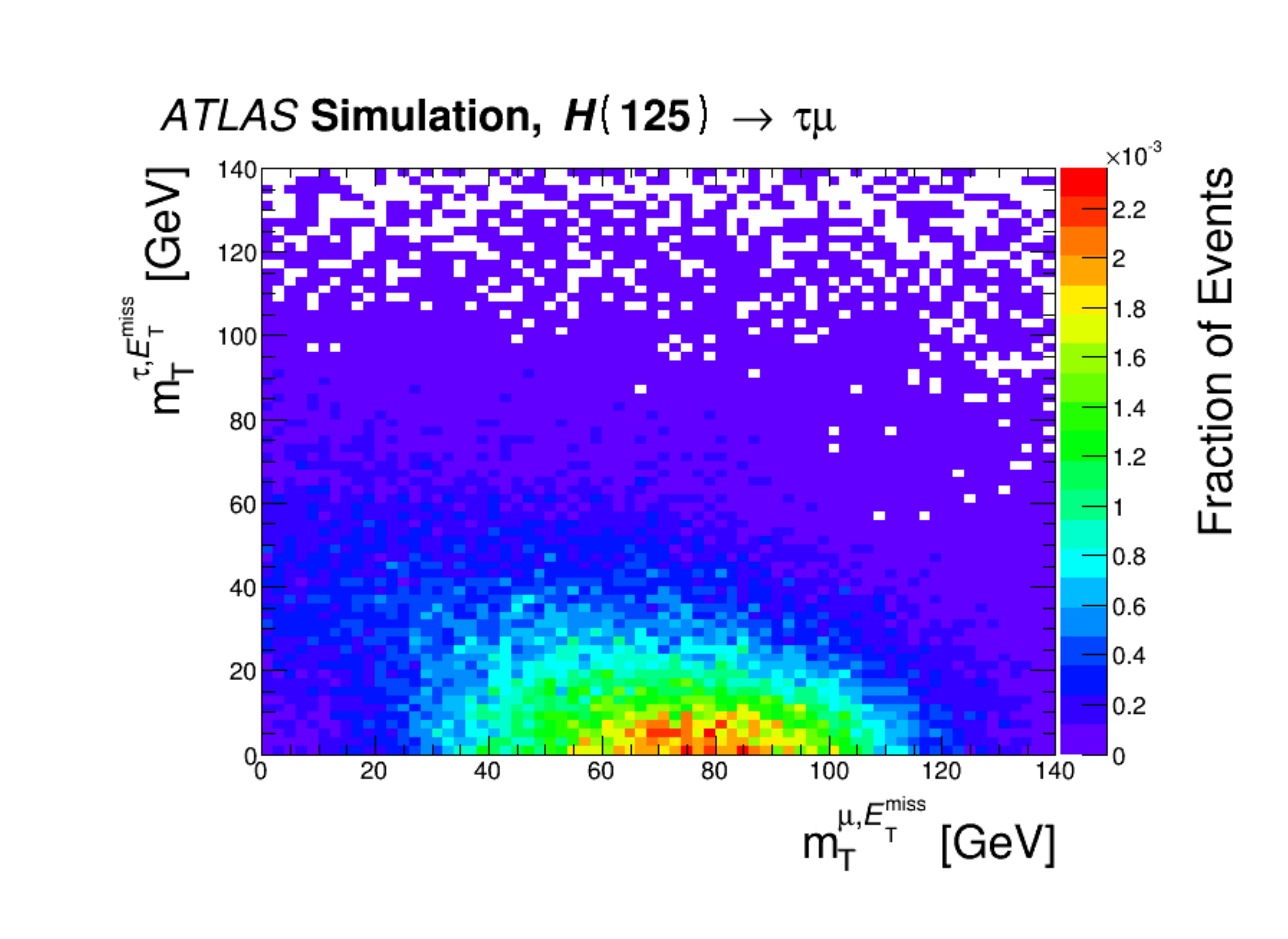}
\includegraphics[width=0.45\textwidth]{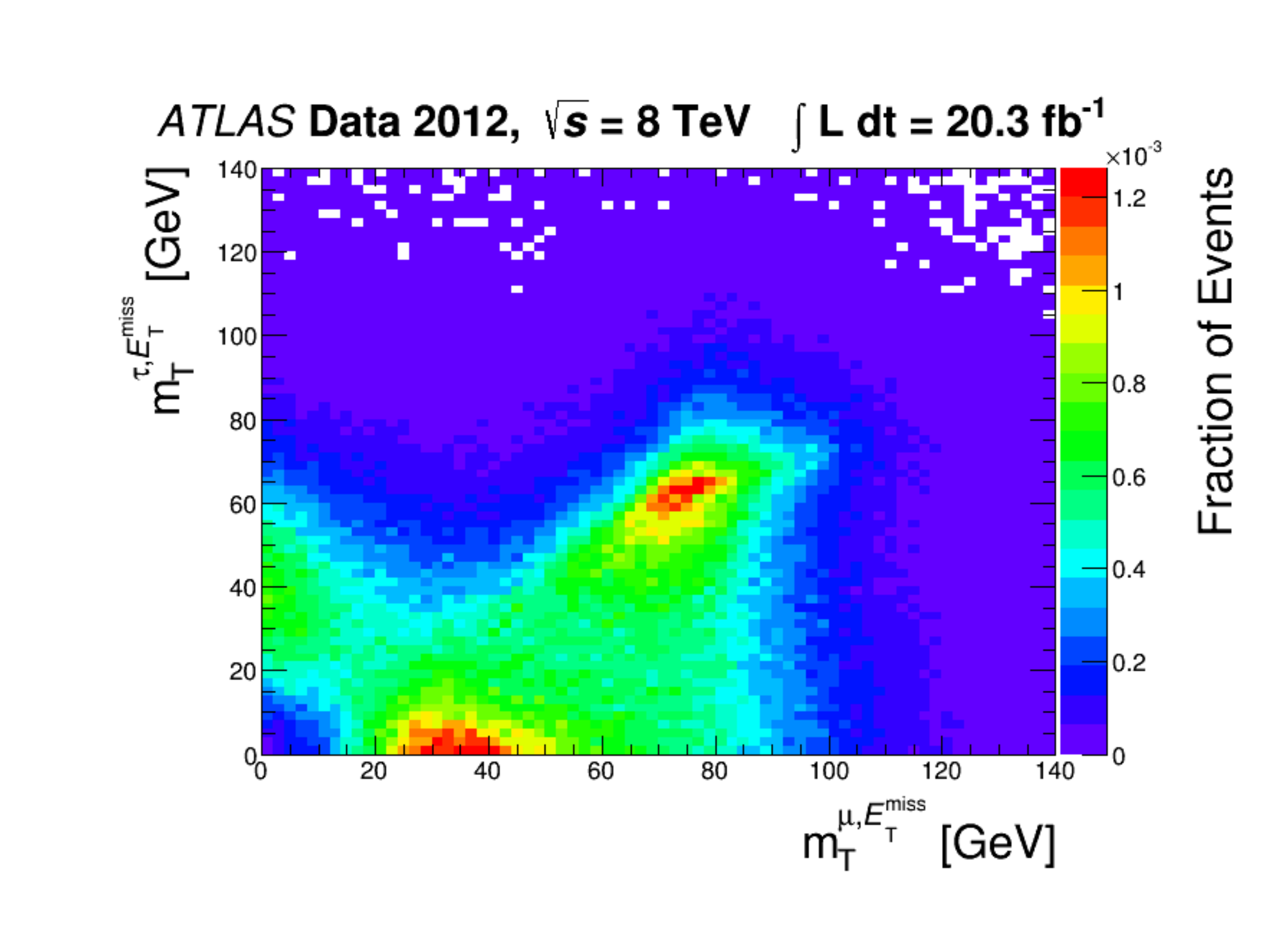}
\caption{Two--dimensional distributions of the transverse mass of the $\mu$--$\met$ system,
$\mT(\mu,\met)$, and that of the $\thad$--$\met$ system, $\mT(\thad,\met)$, in simulated 
$Z/\gamma^{*}\to\tau\tau$ (top left plot), $W$+jets (top right plot), $H\to\mu\tau$ signal 
(bottom left plot) and data (bottom right plot) events. All events are required to have a
well--identified muon and $\thad$ of opposite charge with $\pt(\thad)>20$~GeV and $\pt(\mu)>26$~GeV.}
\label{fig:mt_2Dplots}
\end{figure}

The LFV signal is searched for by performing a fit to the mass distribution in data, \Mmutau, reconstructed 
from the observed muon, $\thad$ and $\met$ objects by means of the Missing Mass Calculator~\cite{mmc} (MMC). 
Conceptually, the MMC is a more sophisticated version of the collinear approximation~\cite{Ellis:1987xu}. 
The main improvement comes from requiring that the relative orientations of the neutrinos and other 
$\tau$--lepton decay products are consistent with the mass and kinematics of a $\tau$--lepton decay. 
This is achieved by maximising a probability defined in the kinematically allowed phase space region.
The MMC used in the $H\to\tau\tau$ analysis~\cite{Aad:2015vsa}
is modified to take into account that there is only one neutrino from a hadronic $\tau$--lepton 
decay in LFV $H\to\mu\tau$ events. For a Higgs boson with $m_{H}=125$~GeV, the reconstructed 
$\Mmutau$ distribution has a roughly Gaussian shape with a full width at half maximum of $\sim$19~GeV.
The analysis is performed ``blinded'' in the 110~GeV$<\Mmutau<$150~GeV regions of SR1 and SR2, which 
contains $\sim$94\% of the expected signal events. 
 
\begin{table}[htb]
  \begin{center}
    \caption{Summary of the event selection criteria used to define
      the signal and various control regions (see text).                         
    \label{tab:cuts}}
    \begin{tabular}{|c|l|l|l|l|}
      \hline
Cut                 & SR1        & SR2        & WCR        & TCR \\  \hline
$\pt(\mu)$         & $>$26~GeV     & $>$26~GeV     & $>$26~GeV     & $>$26~GeV \\ 
$\pt(\thad)$       & $>$45~GeV     & $>$45~GeV     & $>$45~GeV     & $>$45~GeV  \\
$\mT(\mu,\met)$     & $>$40~GeV  & $<$40~GeV  & $>$60~GeV  & --  \\
$\mT(\thad,\met)$   & $<$30~GeV  & $<$60~GeV  & $>$40~GeV  & --   \\
$|\eta(\mu)-\eta(\thad)|$  & $<$2  & $<$2 & $<$2 & $<$2        \\
$N_{\mathrm{jet}}$            &  --  &  --  &  --  & $>$1    \\
$N_{b-\mathrm{jet}}$          &  0  &  0  &  0  & $>$0    \\  \hline
    \end{tabular}
  \end{center}
\end{table}

\section{Background estimation}
\label{sec:backgrounds}

The background estimation method takes into account the background properties and composition discussed 
in section~\ref{sec:selection}. It also relies on the assumption that the shape of the $\Mmutau$ 
distribution for the multi--jet background is the same for OS and SS events. This assumption was verified 
in the published $H\to\tau\tau$ 
search~\cite{HttMoriond2012}. In addition, it was confirmed using a dedicated control region, 
MJCR, with an enhanced contribution from the multi--jet background. Events in this control region are required
to pass all criteria for SR1 and SR2 with the exception of the requirement on $|\eta(\mu)-\eta(\thad)|$, 
which is reversed: $|\eta(\mu)-\eta(\thad)|>2$. Therefore, the number of the total OS background events, 
$N_{\mathrm{OS}}^{\mathrm{bkg}}$ in each bin of the $\Mmutau$ (or any other) distribution in SR1 and SR2 can be 
obtained according to the following formula:
\begin{align}
N_{\mathrm{OS}}^{\mathrm{bkg}} =\rqcd \cdot N_{\mathrm{SS}}^{\mathrm{data}} +N_{\mathrm{OS-SS}}^{Z\to\tau\tau}
+N_{\mathrm{OS-SS}}^{Z\to\mu\mu} +N_{\mathrm{OS-SS}}^{W+\mathrm{jets}} +N_{\mathrm{OS-SS}}^{\mathrm{top}} +N_{\mathrm{OS-SS}}^{VV} +N_{\mathrm{OS-SS}}^{H\to\tau\tau},
\label{eq:bckg_formula}
\end{align}
where the individual terms are described below. $N_{\mathrm{SS}}^{\mathrm{data}}$ is the number of SS data events, 
which are dominated by $W$+jets events but also contain contributions from multi--jet and other backgrounds. The 
fractions of multi--jet background in SS data events inside the 110~GeV$<\Mmutau<$150~GeV mass window are $\sim$17\% 
and $\sim$44\% in SR1 and SR2, respectively. The contributions 
$N_{\mathrm{OS-SS}}^{\mathrm{bkg}-i}=N_{\mathrm{OS}}^{\mathrm{bkg}-i}-\rqcd \cdot N_{\mathrm{SS}}^{\mathrm{bkg}-i}$ are {\it add--on} 
terms for the different backgrounds components (where bkg--$i$ indicates the $i^{\mathrm{th}}$ background source: 
$Z\to\tau\tau$, $Z\to\mu\mu$, $W$+jets, $VV$, $H\to\tau\tau$ and events with $t$--quarks), which also account 
for components of these backgrounds already included in SS data events.\footnote{The $\rqcd \cdot N_{\mathrm{SS}}^{\mathrm{bkg}-i}$
  correction in the {\it add--on} term is needed because same--sign data events include multi--jet as well as 
  electroweak events ($Z\to\tau\tau$, $Z\to\mu\mu$, $W$+jets, $VV$, $H\to\tau\tau$ and events with $t$--quarks) 
  and their contributions cannot be separated.} The factor $\rqcd=N_{\mathrm{OS}}^{\mathrm{multi-jet}}/N_{\mathrm{SS}}^{\mathrm{multi-jet}}$ 
accounts for potential differences in flavour composition (and, as a consequence, in jet~$\to\thad$ fake rates) 
of final--state jets introduced by the same--sign or opposite--sign charge requirements. The value of $\rqcd=\Rqcd$ 
is obtained from a multi--jet--enriched control region in data, as discussed in detail in ref.~\cite{HttMoriond2012}. 
It was verified in the MJCR in this search.    

The largely irreducible $Z/\gamma^{*}\to\tau\tau$ background is modelled by $Z/\gamma^{*}\to\mu\mu$ data events,
where the muon tracks and associated energy deposits in the calorimeters are replaced by the corresponding 
simulated signatures of the final--state particles of the $\tau$--lepton decay. In this approach, essential 
features such as the modelling of the kinematics of the produced boson, the modelling of the hadronic activity 
of the event (jets and underlying event) as well as contributions from pile--up are taken from data. Therefore, 
the dependence on the simulation is minimized and only the $\tau$--lepton decays and the detector response to the 
$\tau$--lepton decay products are based on simulation. This hybrid sample is referred to as embedded data in 
the following. A detailed description of the embedding procedure can be found in ref.~\cite{Aad:2015kxa}. The   
$Z/\gamma^{*}\to\tau\tau$ normalization is a free--floating parameter in the final fit to data and it is mainly 
constrained by events with 60~GeV$<$$\Mmutau$$<$110~GeV in SR2.     

The $W$+jets and $Z\to\mu\mu$ backgrounds are modelled by the ALPGEN~\cite{Mangano:2002ea} event generator interfaced
with PYTHIA8~\cite{Sjostrand:2007gs} to provide the parton showering, hadronization and the modelling of the underlying 
event. In all $W$+jets events, the $\thad$ signature is faked by jets. The WCR region is used to check the modelling 
of the $W$+jets kinematics and to obtain normalizations for OS and SS $W$+jets events. An additional overall normalization 
factor for the $N_{\mathrm{OS-SS}}^{W+\mathrm{jets}}$ term in eq.~(\ref{eq:bckg_formula}) is introduced as a free--floating 
parameter in the final fit in SR1. By studying WCR events and SR1 events with $\Mmutau>150$~GeV (dominated by 
$W$+jets background), it is also found that a $\Mmutau$ shape correction, which depends on the number of jets, 
$\pt(\thad)$ and $|\eta(\mu)-\eta(\thad)|$, needs to be applied in SR1. This correction is derived from SR1 events with 
$\Mmutau>150$~GeV and it is applied to events with all values of $\Mmutau$. A 50\% difference between the SR1--based 
correction and that obtained in WCR is taken as a corresponding modelling uncertainty on the $\Mmutau$ shape for the 
$W$+jets background in SR1. The size of this uncertainty depends on $\Mmutau$ and it is as large as $\pm$10\% for 
$W$+jets events with $\Mmutau<150$~GeV. In the case of SR2, a good modelling of the $N_{\mathrm{jet}}$, $\pt(\thad)$ and
$|\eta(\mu)-\eta(\thad)|$ distributions suggests that such a correction is not needed. 
However, a modelling uncertainty on the $\Mmutau$ shape of the $W$+jets background in SR2 is assigned 
based on the 50\% difference between the default $\Mmutau$ shape and the one obtained after applying the correction 
derived for SR1 events. The size of this uncertainty is below 5\% in the 110~GeV$<\Mmutau<$150~GeV region, that 
contains most of the signal events. It was also checked that applying the same correction in SR2 as in SR1 had 
a negligible effect on the final result and the extracted branching ratio Br($H\to\mu\tau$) (see Section~\ref{sec:results}) 
would only be affected at a level below 3\%. The modelling of jet fragmentation and the underlying 
event has a significant effect on the estimate of the jet~$\to\thad$ fake rate in different regions of the phase 
space and has to be accounted for with a corresponding systematic uncertainty. To estimate this effect, the analysis 
was repeated using a sample of $W$+jets events modelled by ALPGEN interfaced with the HERWIG~\cite{Corcella:2000bw} 
event generator. Differences in the $W$+jets predictions in SR1 and SR2 are found to be $\pm$9\% and $\pm$2\%, 
respectively, and are taken as corresponding systematic uncertainties. 

In the case of the $Z\to\mu\mu$ background, there are two components: events where a muon fakes a $\thad$ and events 
where a jet fakes a $\thad$. Predictions for the shape and normalization of the $Z\to\mu\mu$ (where $\mu\to\thad^{\mathrm{fake}}$) 
background are obtained from simulation. For the $Z\to\mu\mu$ background where a jet is misidentified as a $\thad$ candidate,
the normalization factor and shape corrections, which depend on the number of jets, $\pt(\thad)$ and $|\eta(\mu)-\eta(\thad)|$, 
are derived by using events with two identified OS muons with an invariant mass, $m_{\mu\mu}$, in the range of 80--100~GeV. 
Since this background does not have an OS--SS charge asymmetry, a single correction factor is derived for OS and SS events. 
A 50\% difference between the $\Mmutau$ shape with and without this correction is taken as a corresponding systematic uncertainty.  

The backgrounds with top quarks are modelled by the POWHEG~\cite{Nason:2004rx,Frixione:2007vw,Alioli:2010xd} (for $t\bar{t}$, 
$Wt$ and $s$--channel single--top production) and AcerMC~\cite{Kersevan:2004yg} ($t$--channel single--top production) 
event generators interfaced with PYTHIA8 to provide the parton showering, hadronization and the modelling of the underlying 
event. The TCR is used to check the modelling and to obtain normalizations for OS and SS events with top quarks. The normalization
factors obtained in the TCR are extrapolated into SR1 and SR2, where $t\bar{t}$ and single--top events may have different 
properties. To estimate the uncertainty associated with such an extrapolation, the analysis is repeated using the 
MC@NLO~\cite{Frixione:2002ik} event generator instead of POWHEG for $t\bar{t}$ production.\footnote{The same 
extrapolation uncertainty is assumed for $t\bar{t}$ and single--top backgrounds.} This uncertainty is found to be 
$\pm$7.2\% ($\pm$3.7\%) for backgrounds with top quarks in SR1 (SR2).        

The background due to diboson ($WW$, $ZZ$ and $WZ$) production is estimated from simulation, normalized to 
the cross sections calculated at next--to--leading order (NLO) in QCD~\cite{Campbell:2011bn}. The ALPGEN event generator
interfaced with HERWIG is used to model the $WW$ process, and HERWIG is used for the $ZZ$ and $WZ$ processes.

Finally, events with Higgs bosons produced via gluon fusion or vector--boson fusion (VBF) processes are generated 
at NLO accuracy with POWHEG~\cite{Alioli:2008tz} event generator interfaced with PYTHIA8 to provide the parton 
showering, hadronization and the modelling of the underlying event. The associated production ($ZH$ and $WH$) 
samples are simulated using PYTHIA8. All events with Higgs bosons are produced with a mass of $m_{H}=125$~GeV 
and normalized to cross sections calculated at next--to--next--to--leading order in
QCD~\cite{Anastasiou:2002yz,Ravindran:2003um,Bolzoni:2010xr}. The SM $H\to\tau\tau$ 
decays are simulated by PYTHIA8. The LFV Higgs boson decays are modelled by the EvtGen~\cite{Lange:2001uf} event 
generator according to the phase--space model. In the $H\to\mu\tau$ decays, the $\tau$--lepton decays are treated 
as unpolarised because the left-- and right--handed $\tau$--lepton polarisation states are produced at equal rates. 

All simulated samples are passed through the GEANT4--based ATLAS detector simulation~\cite{Aad:2010ah,Agostinelli:2002hh}. 
The simulated events are overlaid with additional minimum--bias events to account for the effect of multiple $pp$
interactions (pile--up) occurring in the same and neighbouring bunch crossings. 

Figure~\ref{fig:postfit_mmc} shows the $\Mmutau$ distributions for data and the predicted backgrounds in each of 
the signal regions. The backgrounds are estimated using the method described above. The signal efficiencies for 
passing the SR1 or SR2 selection requirements are 2.1\% and 1.5\%, respectively, and the combined efficiency is 3.6\%. 
The numbers of observed events in the data as well as the signal and background predictions in the mass region 
110~GeV$<\Mmutau<$150~GeV can be found in table~\ref{tab:yields}.

\begin{figure}
\centering
\includegraphics[width=0.45\textwidth]{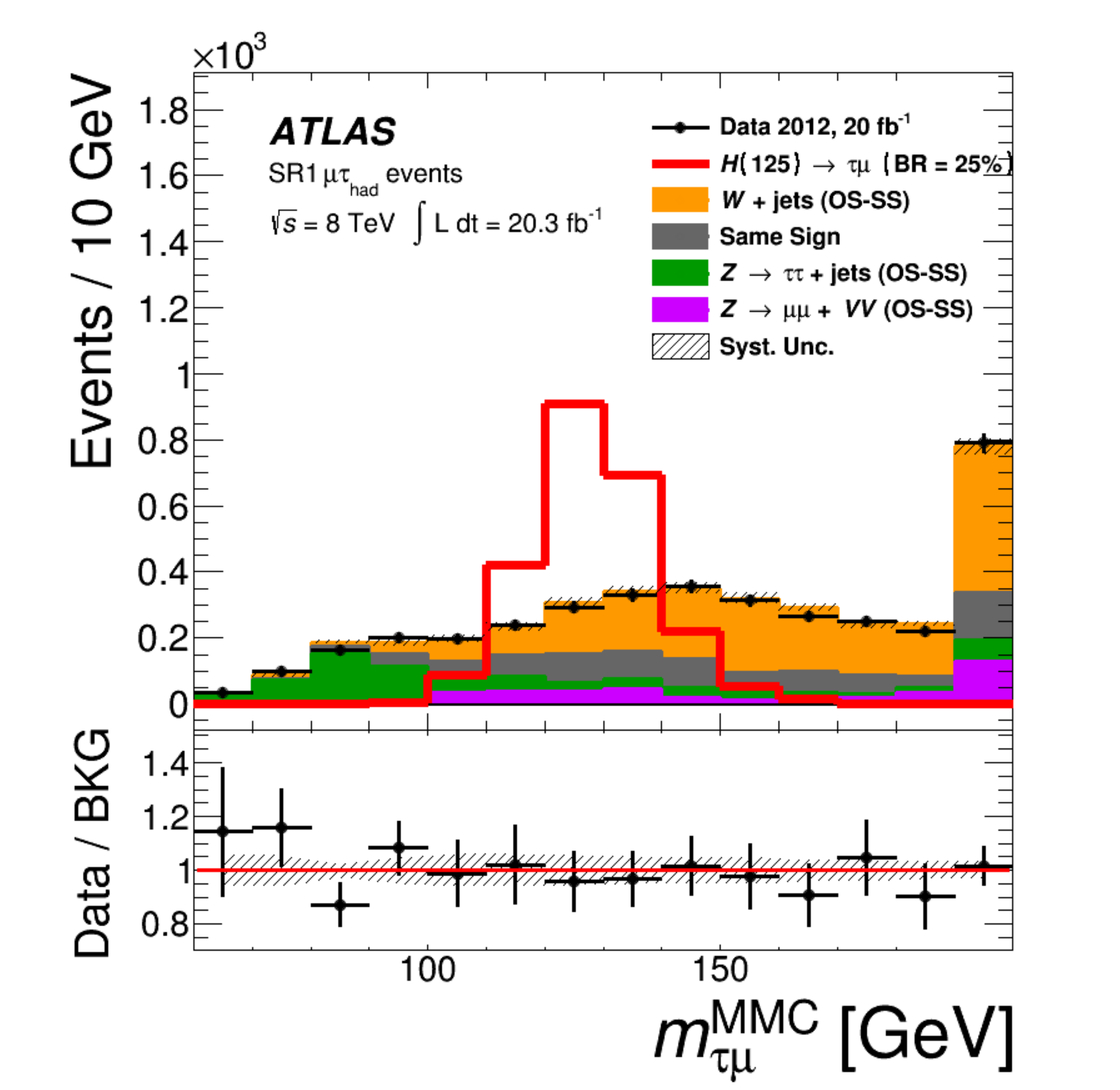}
\includegraphics[width=0.45\textwidth]{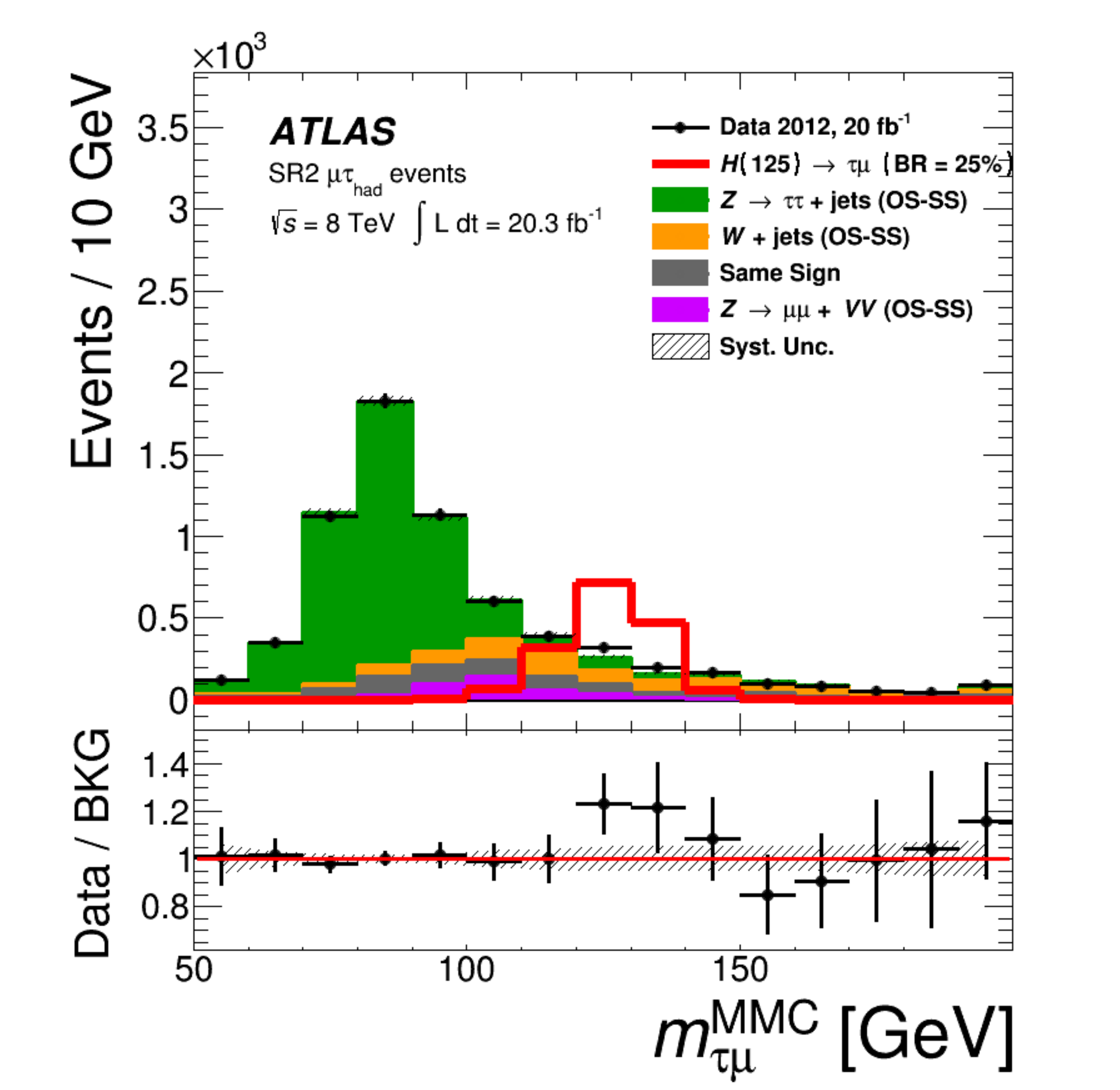}
\caption{Distributions of the mass reconstructed by the Missing Mass
  Calculator, $\Mmutau$, in SR1 (left) and SR2 (right). 
  The background distributions are determined in a global fit.
  The signal distribution corresponds to Br($H\to\mu\tau$)=25\%. The 
  bottom panel of each sub-figure shows the ratio 
  of the observed data and the estimated background. 
  The grey band for the ratio illustrates post--fit systematic 
  uncertainties on the background prediction. The statistical
  uncertainties for data and background predictions are added 
  in quadrature for the ratios. The last bin in each distribution
  contains overflow events.}
\label{fig:postfit_mmc}
\end{figure}

\begin{table}[htb]
  \begin{center}
    \caption{Data yields, signal and post--fit OS--SS background
      predictions (see eq.~(\ref{eq:bckg_formula})) for the 
      110~GeV$<\Mmutau<$150~GeV region. The signal predictions are 
      given for Br($H\to\mu\tau$)=0.77\%. 
      The background predictions are obtained from the combined fit
      to SR1, SR2, WCR and TCR.   
      The post--fit values of systematic uncertainties are 
      provided for the background predictions. 
      For the total background, all correlations between various sources 
      of systematic uncertainties and backgrounds are taken into account.
      The quoted uncertainties represent the statistical (first) and
      systematic (second) uncertainties, respectively.                           
    \label{tab:yields}}
    \begin{tabular}{|c|c|c|}
      \hline
                  & SR1                       & SR2 \\  \hline
Signal            & $69.1 \pm 0.8 \pm 9.2$   & $48.5 \pm 0.8 \pm 7.5$  \\  \hline
$Z\to\tau\tau$    & $133.4 \pm 6.9 \pm 9.1$   & $262.6 \pm 9.7 \pm 18.6$     \\
$W$+jets          & $619 \pm 54 \pm 55$       & $406 \pm 42 \pm 34$        \\
Top               & $39.5 \pm 5.3 \pm 4.7$    & $19.6 \pm 3.1 \pm 3.3$     \\
Same--Sign events & $335 \pm 19 \pm 47$       & $238 \pm 16 \pm 34$        \\
$VV+Z\to\mu\mu$   & $90 \pm 21 \pm 16$        & $81 \pm 22 \pm 17$    \\
$H\to\tau\tau$    & $6.82 \pm 0.21 \pm 0.97$  & $13.7 \pm 0.3 \pm 1.9$   \\
Total background  & $1224 \pm 62 \pm 63$      & $1021 \pm 51 \pm 49$        \\  \hline        
Data              & 1217 & 1075  \\ \hline
    \end{tabular}
  \end{center}
\end{table}

\section{Systematic uncertainties}
\label{sec:systematics}

The largest systematic uncertainties arise from the normalization ($\pm$10\% uncertainty) and modelling\footnote{Some 
of these uncertainties (e.g., uncertainties due to $\Mmutau$ shape corrections and extrapolation uncertainties) are 
discussed in the text above.} of the $W$+jets background. The uncertainties on $\rqcd$ ($\pm$12.7\%) and 
on the normalization ($\pm$6\% uncertainty) and modelling of $Z\to\tau\tau$ also play an important role. The other 
major sources of experimental uncertainty, affecting both the shape and normalization of signal and backgrounds, are 
the uncertainty on the $\thad$ energy scale~\cite{Aad:2014rga} (measured with $\pm$(2--4)\% precision) and 
uncertainties on the embedding method used to model the $Z\to\tau\tau$ background~\cite{Aad:2015vsa}. Less significant 
sources of experimental uncertainty, affecting the shape and normalization of signal and backgrounds, are the uncertainty 
on the jet energy scale~\cite{Aad:2012vm,Aad:2014bia} and resolution~\cite{Aad:2012ag}. The uncertainties
in the $\thad$ energy resolution, the momentum scale and resolution of muons, and the scale uncertainty on $\MET$ due to 
the energy in calorimeter cells not associated with physics objects are taken into account, however, they are found to be 
relatively small. The following experimental uncertainties primarily affect the normalization of signal and 
backgrounds: the $\pm$2.8\% uncertainty on the integrated luminosity~\cite{Aad:2013ucp}, the uncertainty on the $\thad$ 
identification efficiency~\cite{Aad:2014rga}, which is measured to be $\pm$(2--3)\% for 1--prong and $\pm$(3--5)\% for 3--prong 
decays, the $\pm$2.1\% uncertainty for triggering, reconstructing and identifying muons~\cite{Aad:2014rra,Aad:2014sca}, and the $\pm$2\% 
uncertainty on the $b$--jet tagging efficiency~\cite{ATLAS-CONF-2014-046}.      

Theoretical uncertainties are estimated for the signal and for the $H\to\tau\tau$, $VV$ and $Z\to\mu\mu$ (with $\mu\to\thad^{\mathrm{fake}}$) 
backgrounds, which are modelled with the simulation and are not normalized to data in dedicated control regions. Uncertainties 
due to missing higher--order QCD corrections on the production cross sections are found to be~\cite{Dittmaier:2011ti} 
$\pm$10.1$\%$ ($\pm$7.8$\%$) for the Higgs boson production via gluon fusion in SR1 (SR2), $\pm$1$\%$ for the $Z\to\mu\mu$ 
background and for VBF and $VH$ Higgs boson production, and $\pm$5\% for the $VV$ background. The systematic uncertainties 
due to the choice of parton distribution functions used in the simulation are evaluated based on the prescription described 
in ref.~\cite{Dittmaier:2011ti} and the following values are used in this analysis: $\pm$7.5$\%$ for the Higgs boson production 
via gluon fusion, $\pm$2.8$\%$ for the VBF and $VH$ Higgs boson production, and $\pm$4$\%$ for the $Z\to\mu\mu$ and $VV$ 
backgrounds. Finally, an additional $\pm$5.7\% systematic uncertainty on Br($H\to\tau\tau$) is applied to the SM $H\to\tau\tau$ 
background. 

\section{Results}
\label{sec:results}

\begin{figure}
\centering
\includegraphics[width=0.5\textwidth]{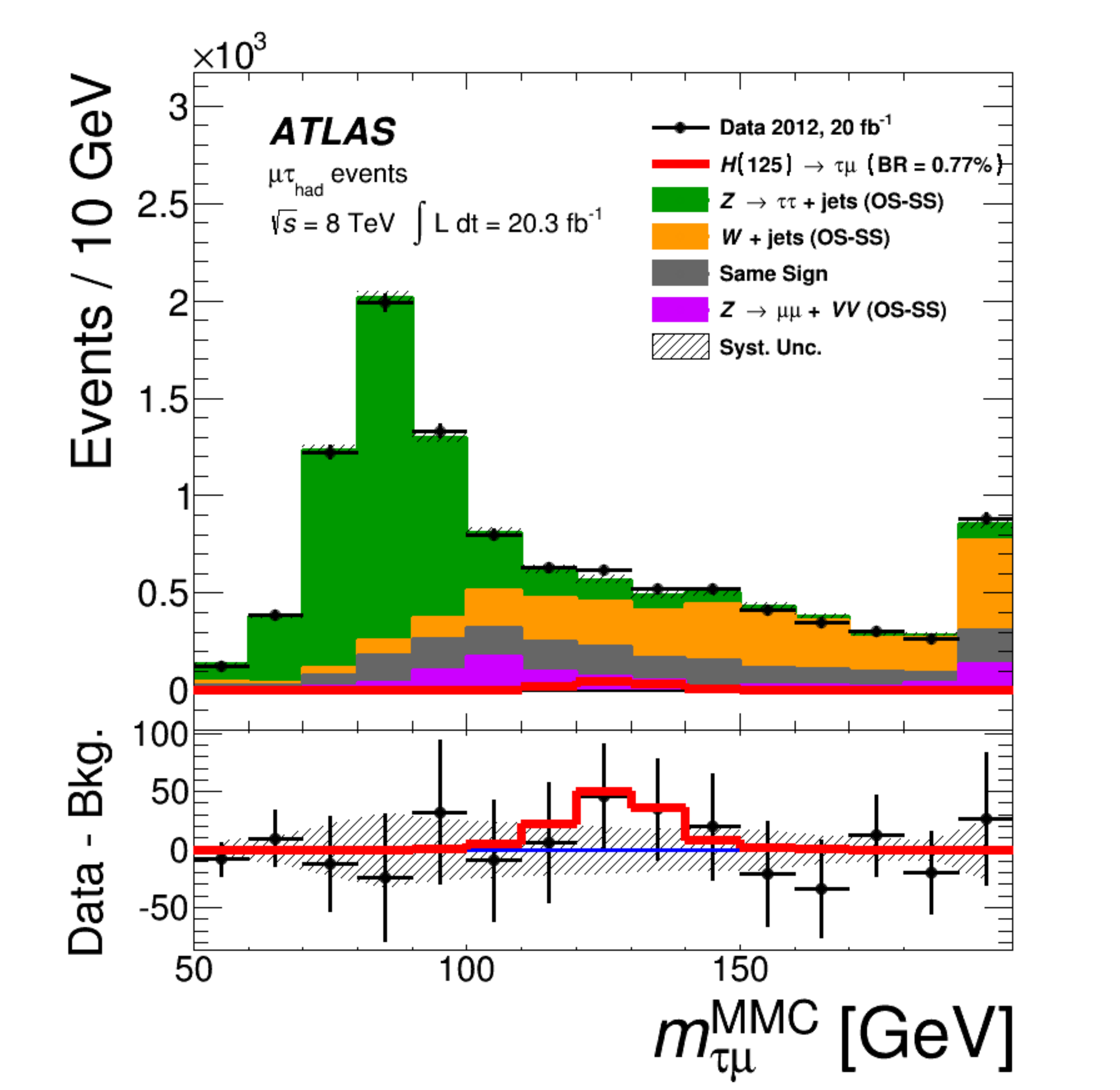}
\caption{ Post--fit combined $\Mmutau$ distribution obtained by adding individual distributions in SR1 
  and SR2. In the lower part of the figure, the data are shown after subtraction of the estimated 
  backgrounds. The grey band in the bottom panel illustrates the post--fit systematic uncertainties on 
  the background prediction. The statistical uncertainties for data and background predictions are added 
  in quadrature on the bottom part of the figure. The signal is shown assuming
  Br($H\to\mu\tau$)=0.77$\%$, the central value of the best fit to
  Br($H\to\mu\tau$). The last bin of the distribution contains overflow events.}
\label{fig:mmc_combined}
\end{figure}

 
A simultaneous binned maximum--likelihood fit is performed on the $\Mmutau$ distributions in SR1 and SR2 and on 
event yields in WCR and TCR to extract the LFV branching ratio Br($H\to\mu\tau$). The fit exploits the control 
regions and the distinct shapes of the $W$+jets and $Z\to\tau\tau$ backgrounds in the signal regions to constrain 
some of the systematic uncertainties. This leads to an improved sensitivity of the analysis. The post--fit $\Mmutau$ 
distributions in SR1 and SR2 are shown in figure~\ref{fig:postfit_mmc}, and the combined $\Mmutau$ distribution 
for both signal regions is presented in figure~\ref{fig:mmc_combined}. Figure~\ref{fig:postfit_mmc} illustrates 
good agreement between data and background expectations in SR1. A small excess of the data over the predicted
background is observed in the 120~GeV$<\Mmutau<$140~GeV region in SR2. This small excess in SR2 has a local significance of 
2.2 standard deviations and a combined significance for both signal regions of 1.3 standard deviations.
This corresponds to a best fit value for the branching fraction of Br($H\to\mu\tau$)=($0.77\pm0.62$)\%. Due to 
the low significance of the observed excess, an upper limit on the LFV branching ratio Br($H\to\mu\tau$) for a 
Higgs boson with $m_{H}=125$~GeV is set using the CL$_s$ modified frequentist formalism~\cite{Read:2002hq} with 
the profile likelihood--ratio test statistics~\cite{Cowan:2010js}. The observed and the median expected 95\% CL 
upper limits are \LimObs\% and \LimExp\%, respectively. Table~\ref{tab:results} provides a summary of 
all results.

\begin{table}[htb]
  \begin{center}
    \caption{The expected and observed 95\% confidence level (CL) upper limits and the best
      fit values for the branching fractions for the two signal regions and their combination.                    
    \label{tab:results}}
    \begin{tabular}{|l|c|c|c|}
      \hline
                                             & SR1                   &    SR2               & Combined \\ \hline
                                            &                       &                      &  \\  
Expected limit on Br($H\to\mu\tau$) [$\%$]  & $1.60^{+0.64}_{-0.45}$  & $1.75^{+0.71}_{-0.49}$ & $1.24^{+0.50}_{-0.35}$ \\  
                                            &                       &                      &  \\  
Observed limit on Br($H\to\mu\tau$) [$\%$]  & 1.55                  & 3.51                 & 1.85  \\  
                                            &                       &                      &  \\ \hline 
                                            &                       &                      &  \\  
Best fit Br($H\to\mu\tau$) [$\%$]           & $-0.07^{+0.81}_{-0.86}$  & $1.94^{+0.92}_{-0.89}$ & 0.77$\pm$0.62   \\ 
                                            &                       &                      &  \\  \hline
    \end{tabular}
  \end{center}
\end{table}

\section{Summary}
\label{sec:summary}

A direct search for lepton--flavour--violating $H\to\mu\tau$ decays of the recently 
discovered Higgs boson is performed in the $\thad$ decay mode of $\tau$--leptons using a data sample of proton--proton 
collisions recorded by the ATLAS detector at the LHC corresponding to an integrated luminosity of 20.3~\ifb\ at a 
centre--of--mass energy of $\sqrt{s}=8$~TeV. The observed and the median expected upper limits at 95\% CL on the 
branching fraction, Br($H\to\mu\tau$), are \LimObs\% and \LimExp\%, respectively. This search places significantly 
more stringent constraints on Br($H\to\mu\tau$) compared to earlier indirect estimates. The result of this analysis 
is found to be consistent with the one published by the CMS Collaboration~\cite{Khachatryan:1994783}.

%% file: Acknowledgements.tex

We thank CERN for the very successful operation of the LHC, as well as the
support staff from our institutions without whom ATLAS could not be
operated efficiently.

We acknowledge the support of ANPCyT, Argentina; YerPhI, Armenia; ARC,
Australia; BMWFW and FWF, Austria; ANAS, Azerbaijan; SSTC, Belarus; CNPq and FAPESP,
Brazil; NSERC, NRC and CFI, Canada; CERN; CONICYT, Chile; CAS, MOST and NSFC,
China; COLCIENCIAS, Colombia; MSMT CR, MPO CR and VSC CR, Czech Republic;
DNRF, DNSRC and Lundbeck Foundation, Denmark; EPLANET, ERC and NSRF, European Union;
IN2P3-CNRS, CEA-DSM/IRFU, France; GNSF, Georgia; BMBF, DFG, HGF, MPG and AvH
Foundation, Germany; GSRT and NSRF, Greece; RGC, Hong Kong SAR, China; ISF, MINERVA, GIF, I-CORE and Benoziyo Center, Israel; INFN, Italy; MEXT and JSPS, Japan; CNRST, Morocco; FOM and NWO, Netherlands; BRF and RCN, Norway; MNiSW and NCN, Poland; GRICES and FCT, Portugal; MNE/IFA, Romania; MES of Russia and NRC KI, Russian Federation; JINR; MSTD,
Serbia; MSSR, Slovakia; ARRS and MIZ\v{S}, Slovenia; DST/NRF, South Africa;
MINECO, Spain; SRC and Wallenberg Foundation, Sweden; SER, SNSF and Cantons of
Bern and Geneva, Switzerland; NSC, Taiwan; TAEK, Turkey; STFC, the Royal
Society and Leverhulme Trust, United Kingdom; DOE and NSF, United States of
America.

The crucial computing support from all WLCG partners is acknowledged
gratefully, in particular from CERN and the ATLAS Tier-1 facilities at
TRIUMF (Canada), NDGF (Denmark, Norway, Sweden), CC-IN2P3 (France),
KIT/GridKA (Germany), INFN-CNAF (Italy), NL-T1 (Netherlands), PIC (Spain),
ASGC (Taiwan), RAL (UK) and BNL (USA) and in the Tier-2 facilities
worldwide.

%% file: atlas_authlist.tex
\begin{flushleft}
{\Large The ATLAS Collaboration}

\bigskip

G.~Aad$^{\rm 85}$,
B.~Abbott$^{\rm 113}$,
J.~Abdallah$^{\rm 151}$,
O.~Abdinov$^{\rm 11}$,
R.~Aben$^{\rm 107}$,
M.~Abolins$^{\rm 90}$,
O.S.~AbouZeid$^{\rm 158}$,
H.~Abramowicz$^{\rm 153}$,
H.~Abreu$^{\rm 152}$,
R.~Abreu$^{\rm 116}$
% ...
\bigskip
\\
% ...
$^{11}$ Institute of Physics, Azerbaijan Academy of Sciences, Baku, Azerbaijan\\
% ...
$^{85}$ CPPM, Aix-Marseille Universit{\'e} and CNRS/IN2P3, Marseille, France\\
% ...
$^{90}$ Department of Physics and Astronomy, Michigan State University, East Lansing MI, United States of America\\
% ...
$^{107}$ Nikhef National Institute for Subatomic Physics and University of Amsterdam, Amsterdam, Netherlands\\
% ...
$^{113}$ Homer L. Dodge Department of Physics and Astronomy, University of Oklahoma, Norman OK, United States of America\\
% ...
$^{116}$ Center for High Energy Physics, University of Oregon, Eugene OR, United States of America\\
% ...
$^{151}$ Institute of Physics, Academia Sinica, Taipei, Taiwan\\
$^{152}$ Department of Physics, Technion: Israel Institute of Technology, Haifa, Israel\\
$^{153}$ Raymond and Beverly Sackler School of Physics and Astronomy, Tel Aviv University, Tel Aviv, Israel\\
% ...
$^{158}$ Department of Physics, University of Toronto, Toronto ON, Canada\\
% ...
\end{flushleft}
